\documentclass[aps,prl,10pt,twocolumn,nofootinbib,groupedaddress,floatfix]{revtex4-2}

\usepackage[colorlinks=true,urlcolor=blue,linkcolor=blue,citecolor=blue]{hyperref}
\usepackage{amsmath,amssymb,amstext,amsbsy,amsfonts,amsthm,graphicx,microtype,booktabs}
\usepackage[capitalize]{cleveref}
\usepackage{macros}
\renewcommand{\section}[1]{\phantomsection\addcontentsline{toc}{section}{#1}\textit{#1}---\unskip\ignorespaces}

\usepackage{fontawesome5}
\makeatletter
\newcommand{\amarki}{\color{Blue}{\faCoffee}}
\newcommand{\amarkii}{\color{Blue}{\faBeer}}
\newcommand{\amarkiii}{\color{Blue}{\faRebel}}
\newcommand{\amarkiv}{\color{Blue}{\faPaperPlane[regular]}}
\def\@fnsymbol#1{{\ifcase#1\or \amarki\or \amarkii\or \amarkiii\or \amarkiv \else\@ctrerr\fi}}
\makeatother

\begin{document}

\title{Detectable and defect-free dark photon dark matter}

\author{David Cyncynates}
\email{davidcyn@uw.edu}
\affiliation{Department of Physics, University of Washington, Seattle, WA 98195, U.S.A.}

\author{Zachary J. Weiner}
\email{zweiner@uw.edu}
\affiliation{Department of Physics, University of Washington, Seattle, WA 98195, U.S.A.}

\begin{abstract}
Ultralight dark photons are compelling dark matter candidates, but their allowed kinetic mixing with the Standard Model photon
is severely constrained by requiring that the dark photons do not collapse into a cosmic string network in the early Universe.
Direct detection in minimal production scenarios for dark photon dark matter is strongly limited, if not entirely excluded; discovery of sub-meV dark photon dark matter would therefore point to a nonminimal dark sector.
We describe a model that evades such constraints, capable of producing cold dark photons in any parameter space accessible to future direct detection experiments.
The associated production dynamics yield additional signatures in cosmology and small-scale structure, allowing for possible positive identification of this particular class of production mechanisms.
\end{abstract}

\maketitle

% \section{Introduction}
Evidence for cold dark matter abounds in astrophysical and cosmological observations~\cite{Bertone:2004pz,Bertone:2016nfn,Buckley:2017ijx}, but not for
its fundamental nature---the mass and spin of its constituents or its interactions with the visible
sector.
Dark photons are among the best-motivated candidates for new light degrees of freedom and are common features of grand unified theories and string
theory~\cite{Beasley:2008dc,Abel:2008ai,Donagi:2008kj,Blumenhagen:2008zz,Arvanitaki:2009fg,Blumenhagen:2009gk,Goodsell:2009xc,Bullimore:2010aj}.
They may constitute all the dark matter in scenarios ranging from minimal gravitational production
during inflation~\cite{Graham:2015rva,Ema:2019yrd,Ahmed:2020fhc,Kolb:2020fwh,Nakai:2020cfw} to
nonthermal mechanisms involving additional new degrees of
freedom~\cite{Agrawal:2018vin,Bastero-Gil:2018uel,Co:2018lka,Dror:2018pdh,Co:2021rhi,Adshead:2023qiw}.
At low energies, a dark photon can interact with the Standard Model (SM) through kinetic mixing with
the ordinary photon, yielding signatures in cosmology~\cite{Arias:2012az, Foot:2014uba,
McDermott:2019lch, Caputo:2020rnx, Caputo:2020bdy, Witte:2020rvb, Chluba:2024wui, East:2022rsi},
astrophysics~\cite{Redondo:2008aa, Zechlin:2008tj, Foot:2014osa, Dubovsky:2015cca, Vinyoles:2015aba,
Foot:2016wvj, Baryakhtar:2017ngi, Hong:2020bxo, Wadekar:2019mpc, Bi:2020ths, Fedderke:2021aqo,
East:2022ppo, East:2022rsi, Li:2023vpv, Amin:2023imi}, and the laboratory~\cite{Redondo:2008aa,
Ehret:2010mh, Jaeckel:2010xx, Povey:2010hs, Bahre:2013ywa, Betz:2013dza, Inada:2013tx,
Parker:2013fxa, Schwarz:2015lqa, Arias:2016vxn, Kroff:2020zhp, Romanenko:2023irv}; numerous dark
matter haloscopes are poised to probe a vast space of unexplored dark photon masses and kinetic
mixing~\cite{Suzuki:2015sza, Knirck:2018ojz, Brun:2019kak, Hochberg:2019cyy, Nguyen:2019xuh,
Phipps:2019cqy, SuperCDMS:2019jxx, XENON:2019gfn, An:2020bxd, Dixit:2020ymh, FUNKExperiment:2020ofv,
SENSEI:2020dpa, Tomita:2020usq, XENON:2020rca, Chiles:2021gxk, Fedderke:2021rrm, Manenti:2021whp,
An:2022hhb, Cervantes:2022yzp, DarkSide:2022knj, DOSUE-RR:2022ise, Fan:2022uwu, Ramanathan:2022egk,
Bajjali:2023uis, Gelmini:2020kcu, BREAD:2021tpx, Godfrey:2021tvs, Baryakhtar:2018doz, LZ:2021xov,
Gelmini:2020kcu, Chaudhuri:2014dla}.
Such theoretical and observational promise demands understanding whether the parameter space within experimental reach also allows for consistent dark photon dark matter production.

Recent work demonstrated a stringent upper bound on the kinetic mixing that allows for viable dark
photon dark matter~\cite{East:2022rsi}: the dark photon backreacts on the Higgs responsible for its
mass and, with large enough couplings at large enough energy density, can restore the dark
$\mathrm{U}(1)_D$ gauge symmetry.
The associated Goldstone boson winds about sites of symmetry restoration, seeding
string vortices that deplete the energy in the cold, coherent dark electromagnetic fields.
Such a defect network dilutes like radiation and cannot be the dark matter.

\begin{figure}
    \centering
    \includegraphics[width=\columnwidth]{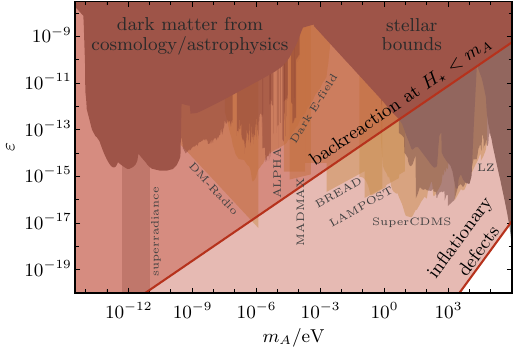}
    \caption{
        Dark photon dark matter parameter space, including exclusions from astrophysical~\cite{Redondo:2008aa,Zechlin:2008tj,Dubovsky:2015cca,Vinyoles:2015aba,Baryakhtar:2017ngi,Hong:2020bxo,Wadekar:2019mpc,Bi:2020ths,Fedderke:2021aqo,East:2022ppo,East:2022rsi,Li:2023vpv}, cosmological~\cite{Arias:2012az,McDermott:2019lch,Caputo:2020rnx,Caputo:2020bdy,Witte:2020rvb,Chluba:2024wui,Arsenadze:2024ywr,East:2022rsi}, and haloscope observations~\cite{Suzuki:2015sza,Knirck:2018ojz,Brun:2019kak,Hochberg:2019cyy,Nguyen:2019xuh,Phipps:2019cqy,SuperCDMS:2019jxx,XENON:2019gfn,An:2020bxd,Dixit:2020ymh,FUNKExperiment:2020ofv,SENSEI:2020dpa,Tomita:2020usq,XENON:2020rca,Chiles:2021gxk,Fedderke:2021rrm,Manenti:2021whp,An:2022hhb,Cervantes:2022yzp,DarkSide:2022knj,DOSUE-RR:2022ise,Fan:2022uwu,Ramanathan:2022egk,Bajjali:2023uis} in gray, and experimental prospects~\cite{Gelmini:2020kcu,BREAD:2021tpx,Godfrey:2021tvs,Baryakhtar:2018doz,LZ:2021xov,Gelmini:2020kcu,Chaudhuri:2014dla} in yellow.
        For inflationary production~\cite{Graham:2015rva}, defect formation with a fiducial kinetic mixing
        $\varepsilon \sim e \gD / 16\pi^2$~\cite{Holdom:1985ag,Rizzo:2018vlb} excludes
        the bulk of parameter space~\cite{East:2022rsi} (red shaded region).
        Most other models effectively produce dark photons no later than when the Hubble rate
        $H_\star \approx \mA$, bounding the kinetic mixing to be below the upper red line.
    }
    \label{fig:existing-parameter-space}
\end{figure}

The dark gauge coupling $\gD$, which controls the strength of backreaction of dark gauge bosons onto the Higgs,
is a free parameter in all production mechanisms and can simply be tuned small enough to
avoid defect formation.
But the dark photon's kinetic mixing with the Standard Model photon $\varepsilon$ is generated by heavy fermions charged under both $\mathrm{U}(1)_Y$ and $\mathrm{U}(1)_D$ and is therefore also proportional to the dark gauge coupling $\gD$~\cite{Holdom:1985ag}.
\cref{fig:existing-parameter-space} shows that, for known production mechanisms,
the prospects for probing the kinetic mixing of dark photon dark matter are severely limited, if not entirely absent.
Direct detection in most parameter space would therefore point to a nonminimal dark sector.

In this letter we describe an extension of the Abelian-Higgs model that realizes cold dark photon dark matter with
kinetic mixing detectable by any planned or proposed laboratory experiment.
We discuss additional signatures in cosmology and small-scale structure that could corroborate the nonminimality of the model.
A companion article~\cite{Cyncynates:2024yxm} explores further generalizations thereof, discusses
the implications of defect formation on existing dark photon production mechanisms in detail, and
studies nondynamical, \textit{ad hoc} means to generate a hierarchy between the kinetic mixing and
the dark gauge coupling (namely, the clockwork mechanism~\cite{Giudice:2016yja}).

% \section{Dynamics}
The Abelian-Higgs theory is described by the Lagrangian\footnote{
    We use natural units in which $\hbar = c = 1$ and the reduced Planck mass
    $\Mpl = 1/\sqrt{8\pi G}$, fix a cosmic-time Friedmann-Lema\^itre-Robertson-Walker (FLRW) metric
    $\mathrm{d} s^2 = \mathrm{d} t^2 - a(t)^2 \delta_{i j} \mathrm{d} x^i \mathrm{d} x^j$
    with $a(t)$ the scale factor, and employ the Einstein summation convention for spacetime indices.
    Dots denote derivatives with respect to cosmic time $t$, and the Hubble rate is
    $H \equiv \dot{a} / a$.
}
\begin{align}
    \label{eqn:abelian-higgs-action-unbroken}
    \mathcal{L}_\mathrm{AH}
    &=
            - \frac{1}{4}\Fp_{\mu\nu} \Fp^{\mu\nu}
            + \frac{1}{2} D_\mu \Phi \left( D^\mu \Phi \right)^\ast- V_\Phi(\Phi),
\end{align}
where $\Ap$ is the dark photon, $\Phi$ is the dark Higgs field,
$D_\mu = \partial_\mu  - \I \gD \Ap_\mu$ is the gauge covariant derivative, and the Higgs potential
$V_\Phi$ has the usual symmetry breaking form,
\begin{align}
    V_\Phi(\Phi)
    &= \frac{\lambda}{4} \left( \abs{\Phi}^2 - \vev^2 \right)^2.
\end{align}
In the broken phase, the dark photon acquires a mass $\mA = \gD\vev$ and contributes
$\gD^2 \vert \Phi \vert^2 \Ap_\mu \Ap^\mu / 2$ to the Higgs's effective potential.
If this contribution (which coincides with the dark photon's energy density $\rho_\Ap$ when it is
nonrelativistic) exceeds $\lambda v^4$, then the dark photon backreacts strongly onto the Higgs and
generally seeds topological defects~\cite{East:2022rsi}.
If dark photon dark matter is produced at a Hubble rate $H_\star$,
at which point
$\Ap_\mu \Ap^\mu \sim \rho_\Ap / \mA^2 \sim 3 H_\mathrm{eq}^2 \Mpl^2 (H_\star / H_\mathrm{eq})^{3/2} / 2$,
evading strong backreaction onto the dark Higgs requires\footnote{
    This threshold assumes the dark photon is composed of nonrelativistic modes; we discuss its
    generalization in Ref.~\cite{Cyncynates:2024yxm}.
    \Cref{eqn:vortex-formation-threshold} also expressed redshifting factors in terms of the Hubble
    parameter by assuming a radiation-dominated Universe.
}
\begin{align}
    \label{eqn:vortex-formation-threshold}
    \gD
    &\lesssim 10^{-14}
        \lambda^{1/4}
        \left( \frac{\mA}{\ueV} \right)^{5/8}
        \left( \frac{H_{\star}}{\mA} \right)^{-3/8}.
\end{align}
In the minimal setup, kinetic mixing is generated by loops of a few heavy fermions with
$\mathcal{O}(1)$ charge~\cite{Holdom:1985ag} with $\varepsilon \sim g_D e / 16 \pi^2$, so this
constraint strongly limits direct detection prospects, as illustrated in
\cref{fig:existing-parameter-space}.
The general considerations leading to \cref{eqn:vortex-formation-threshold} motivate two possible
solutions: either modulate the parameters of the Abelian-Higgs theory to raise the threshold for
defect formation or delay production as late as possible, such that the dark photon never has enough
energy density to exceed the threshold.

These possibilities may be realized by extending the Abelian-Higgs theory
\cref{eqn:abelian-higgs-action-unbroken} with couplings to a singlet scalar $\phi$ as
\begin{align}\label{eqn:scalar-abelian-higgs-action-unbroken}
\begin{split}
    \mathcal{L}
    &= - \frac{W(\phi)}{4} F_{\mu\nu} F^{\mu\nu}
        + \frac{X(\phi)}{2} D_\mu \Phi \left( D^\mu \Phi \right)^\ast
    \\ &\hphantom{ {}={} }
        + Y(\phi) V_\Phi(\Phi)
        + \frac{1}{2} \partial_\mu \phi \partial^\mu \phi
        - V(\phi).
\end{split}
\end{align}
We discuss concrete choices of the coupling functions $W$, $X$, and $Y$ below.
The dark Higgs and photon are made canonical via the rescalings
$\Psi = \sqrt{X(\phi)} \Phi$ and $\mathcal{A}_\mu = \sqrt{W(\phi)} A_\mu$.
Written in terms of the canonical fields, the Higgs's potential is
\begin{align}
    Y(\phi) V_\Phi(\Phi)
    &= \frac{\lambda Y(\phi)}{4 X(\phi)^2} \left( \abs{\Psi}^2 - X(\phi) v^2 \right)^2,
    \label{eqn:higgs-potential-canonical}
\end{align}
and its covariant derivative is
\begin{align}
    \sqrt{X(\phi)} D_\mu \Phi
    &= \partial_\mu \Psi
        - \frac{\I g_D \mathcal{A}_\mu \Psi}{\sqrt{W(\phi)}}
        - \frac{\partial_\mu \sqrt{X(\phi)}}{\sqrt{X(\phi)}} \Psi.
    \label{eqn:higgs-covariant-derivative-canonical}
\end{align}
The form of \cref{eqn:higgs-potential-canonical,eqn:higgs-covariant-derivative-canonical} motivates
absorbing the $\phi$ dependence of the theory into its fundamental parameters as
$g_D(\phi) \equiv g_D / \sqrt{W(\phi)}$,
$\lambda(\phi) \equiv \lambda Y(\phi) / X(\phi)^2$, and
$v(\phi) \equiv v \sqrt{X(\phi)}$.
If the scalar is homogeneous, i.e., $\phi(t, \mathbf{x}) = \bar{\phi}(t)$, then its cosmological
evolution permits the theory's parameters to vary over cosmological history.

The threshold for defect formation then depends on the scalar as
$\lambda(\phi) v(\phi)^4 = \lambda v^4 Y(\phi)$.
One might be tempted to arrange for $Y(\phi) \gg 1$ to simply raise the threshold for defect formation as high as needed.
This solution is no different than simply taking large $\lambda$ in the bare Abelian-Higgs Lagrangian, and it requires the Higgs to be a composite degree of freedom because fundamental Higgs scattering violates perturbative unitarity for $\lambda \gtrsim 4 \pi$.
{}  % without this, random words in the preceding sentence don't show up in the pdf....
Such issues aside, a still more appealing solution would utilize the scalar field's dynamics to not only prevent defect formation but also generate the dark photon relic abundance.

\begin{figure}
    \centering
    \includegraphics[width = \columnwidth]{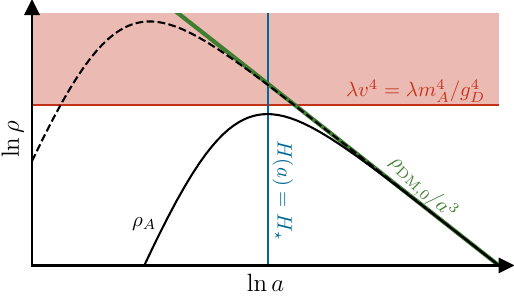}
    \caption{
        Illustration of dark photon production that avoids defect formation by delaying the time of production.
        After it is produced and becomes nonrelativistic, the dark matter has a known energy density at any scale factor $a$ (indicated by the green line)
        extrapolated from its present-day value.
        For any choice of model parameters, its energy density would exceed
        the threshold for defect formation (in the red shaded region) at some early time.
        By sufficiently delaying the production of dark photons, i.e., until some critical Hubble rate $H_\star$ (indicated by the blue line), they never trigger defect formation (per the solid black curve).
        Dark photons produced too early (as in the dashed black curve) collapse into a string network that is not viable cold dark matter.
    }
    \label{fig:schematic}
\end{figure}

We turn to dynamical mechanisms that evade defect formation by delaying production, illustrated by \cref{fig:schematic}.
In general, the mass sets a kinematic barrier for particle production; resonant production with rolling scalars effectively requires the scalar's mass $m_\phi \gtrsim \mA$ for efficient dark photon production.
As the scalar starts rolling when $H\sim m_\phi \gtrsim \mA$, in these scenarios production typically occurs no later than $H\sim \mA$.
On the other hand, scalar couplings offer a means to suppress the dark photon's mass in the early Universe, since $\mA(\phi) \equiv g_D(\phi) v(\phi) = \mA \sqrt{X(\phi) / W(\phi)}$.
In such a scenario, $X(\phi)$ and/or $W(\phi)$ must evolve so that the theory's parameters take on their bare values at the present day.
However, the scalar couplings generate derivative interactions that cannot necessarily be neglected as $\bar{\phi}$ evolves.
This feature is precisely what enables production of a relic abundance of dark photons via tachyonic resonance, familiar from other dark photon models~\cite{Agrawal:2018vin,Bastero-Gil:2018uel,Co:2018lka,Dror:2018pdh,Co:2021rhi}.

In the presence of a homogeneous scalar, the linearized equation of motion for the transverse polarizations of the dark photon $\mathcal{A}_\pm$ is
\begin{align}
    0
    &= \ddot{\mathcal{A}}_\pm
        + H \dot{\mathcal{A}}_\pm
        + \omega_\pm^2 \mathcal{A}_\pm,
    \label{eqn:transverse-eom-rescaled}
\end{align}
with an effective squared frequency
\begin{align}
    \omega_\pm^2
    &= \frac{k^2}{a^{2}}
        + \mA^2\frac{\bar X}{\bar W}
        - \frac{H}{2}
        \frac{\dot{\bar{W}}}{\bar{W}}
        - \frac{\partial_t^2 \sqrt{\bar{W}}}{\sqrt{\bar{W}}}
    \label{eqn:transverse-omega}
\end{align}
[with the shorthand $\bar{W} = W(\bar{\phi})$ and $\bar{X} = X(\bar{\phi})$].
Whenever $\omega_\pm^2$ is negative (i.e., due to the coupling terms outweighing the mass and
momentum contributions), the transverse dark photon modes grow exponentially.
The kinematic requirement that $W$ decreases from the large values that suppress $\mA$ at early
times enables this tachyonic resonance (which is more easily seen from the noncanonical field
$A_\pm$'s friction term $\sim \dot{\bar{W}} \dot{A}_\pm$; see Ref.~\cite{Cyncynates:2024yxm} for
more exposition).
Since the transverse modes are derivatively coupled only to the kinetic function $W(\phi)$,
we set $X(\phi) = 1$ and $Y(\phi) = 1$ from here on out.\footnote{
    The longitudinal mode has derivative couplings via $X(\phi)$, though efficient production requires
    that $X$ decreases with time; we discuss this and other possibilities in Ref.~\cite{Cyncynates:2024yxm}.
}

To illustrate the production mechanism, we present a concrete example and compute the dark photon
relic abundance, present the expanded parameter space accessible to direct detection experiments,
and discuss other constraints on the model.
We consider a so-called runaway potential~\cite{Damour:2002nv,Gasperini:2001pc,Damour:2002mi} for the scalar where
\begin{align}
    V(\phi)
    &= M^2 f^2 e^{- \phi / f}
    \equiv m_\phi^2 f^2 e^{- \left( \phi - \phi_0 \right) / f}.
\end{align}
The latter equality defines $m_\phi$ as the scalar's effective mass at its homogeneous initial
condition $\bar{\phi} = \phi_0$; the scalar remains frozen until $H \approx m_\phi$.
An approximate solution to the scalar's homogeneous equation of motion,
$0 = \ddot{\bar{\phi}} + 3 H \dot{\bar{\phi}} + V'(\bar{\phi})$, is
\begin{align}
    \bar{\phi}(t)
    &= \phi_0 + f \ln \left[ 1 + (m_\phi t)^2 \right].
    \label{eqn:phi-sol}
\end{align}
Full solutions exhibit moderate oscillations in $\ln m_\phi t$ about \cref{eqn:phi-sol}.
Without loss of generality, we take $\phi_0 < 0$ and assume a coupling of the form
\begin{align}
    W(\phi)
    &= 1 + e^{- \kincoup \phi / f},
    \label{eqn:coupling-function-1-plus-exp}
\end{align}
constructed such that $W \approx 1$ after $\phi$ crosses zero, which occurs at
$m_\phi t_\star \approx e^{- \phi_0 / 2 f}$ when the Hubble rate is
$H_\star = 1 / 2 t_\star = M / 2$.

\begin{figure}
    \centering
    \includegraphics[width=\columnwidth]{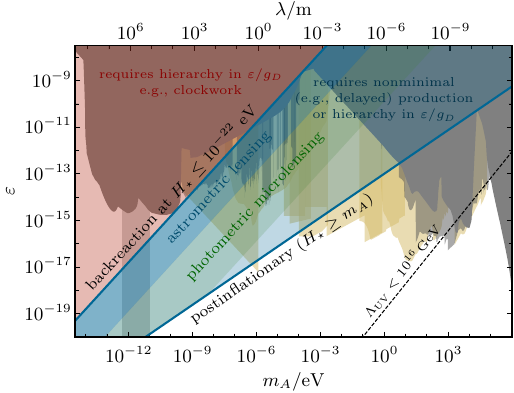}
    \caption{
        Parameter space available for dark photon dark matter produced by a kinetically coupled
        scalar.
        Requiring negligible backreaction at observed epochs
        [\cref{eqn:vortex-formation-threshold} with $H_\star = 10^{-22}~\mathrm{eV}$]
        excludes the red parameter space for the fiducial expectation
        $\varepsilon \sim \gD$, independent of the production mechanism.
        Nonthermal production occurring when the Hubble rate is near the dark photon's mass, e.g.,
        via axion or scalar oscillations~\cite{Agrawal:2018vin, Bastero-Gil:2018uel, Co:2018lka,
        Dror:2018pdh, Co:2021rhi, Adshead:2023qiw}, is viable below the blue ``postinflationary''
        line, while the region between the blue lines is opened up in the kinetically coupled scalar
        we consider by delaying dark photon production to parametrically later times.
        The shaded dark blue and green regions indicate where enhanced DM substructure at $k_\star$
        [\cref{eqn:kpeak}] could be probed by future astrometric~\cite{VanTilburg:2018ykj} and
        photometric~\cite{Arvanitaki:2019rax, Dai:2019lud} surveys, respectively.
        Assuming weak gravity conjectures are true and apply at the displaced field values required
        by the kinetic coupling, the dashed black line indicates the maximum achievable gauge
        coupling for which local quantum field theory holds during, e.g., high-scale inflation.
        The top axis depicts the Compton wavelength of dark photons with a given mass,
        $\lambda = 2 \pi / \mA$.
    }
    \label{fig:kinetic-parameter-space}
\end{figure}

Reference~\cite{Cyncynates:2024yxm} derives analytic approximations to the dark photon dynamics and
relic abundance from the tachyonic instability sourced by the runaway scalar, finding
\begin{align}
    \frac{\Omega_\Ap}{\Omega_\mathrm{DM}}
    &\approx \frac{2}{3}
        \mathcal{N}_{\kincoup} \left( \frac{m_\phi}{2H_\star} \right)^{2\kincoup-1}
        \left( \frac{2H_\star}{\mA} \right)^{\frac{\kincoup - 7/2}{\kincoup + 1}}
        \frac{H_\star^2}{\Mpl^2}
        \sqrt{\frac{H_\star}{H_\mathrm{eq}}}
    ,
    \label{eqn:relic-abundance}
\end{align}
where $H_\mathrm{eq} \approx 2.26 \times 10^{-28}~\mathrm{eV}$ is the Hubble rate at
matter-radiation equality and $\mathcal{N}_\kincoup$ a coefficient whose full expression appears in
Ref.~\cite{Cyncynates:2024yxm} and scales with $\kincoup^{13/2}$ at
large $\kincoup$.
Production may occur as late as desired by choosing $m_\phi$ or $\phi_0$ to set $t_\star$.
Achieving the correct relic abundance requires $- \kincoup \phi_0 / f$ in the range of $150 - 250$; the
dark photon's initial mass $\mA(\phi_0) \approx \mA e^{\kincoup \phi_0 / 2}$ is then more than small
enough (compared to $m_\phi$) for efficient production.
For cosmology to proceed as observed, the dark matter must exist by, say, a redshift $z \sim 10^6$
or so, making $H_\star = 10^{-22}~\mathrm{eV}$ a useful benchmark.
As evinced by \cref{fig:kinetic-parameter-space}, such late production is sufficient for viable dark
photon dark matter in reach of any future experiment.

Other resonant production mechanisms (via oscillating pseudoscalars~\cite{Agrawal:2018vin,Bastero-Gil:2018uel,Co:2018lka,Dror:2018pdh,Co:2021rhi} or scalars~\cite{Adshead:2023qiw}) require the system to reach a regime where the dark photon backreacts on the scalar sourcing it---otherwise, the dark matter would mostly comprise scalars rather than vectors.
These nonlinear dynamics can only be understood with 3D simulations, and the energy exchange that occurs at backreaction often results in a rough equipartition between the dark photon and scalar.
An attractive feature of scenarios involving runaway scalars is that they become energetically subdominant without relying on nonlinear dynamics.
The runaway scalar solution uniquely tracks the background such that its relative abundance is
\begin{align}
    \Omega_\phi(t)
    &\equiv \frac{\rho_\phi(t)}{3 H(t)^2 \Mpl^2}
    \approx \left( \frac{2 f}{\Mpl} \right)^2
\end{align}
in both the radiation- and matter-dominated epochs~\cite{Copeland:1997et, Steinhardt:1999nw,
Copeland:2006wr}.
Comparing to the abundance of the dark photons $\Omega_{\Ap} = \rho_{\Ap} / 3 H^2 \Mpl^2$ provides a
reasonable proxy to assess whether backreaction is important.
At production (when the two decouple), the dark photons have an abundance
$\Omega_{\gamma'}(t_\star) \sim \left[ H_\mathrm{eq} / H_\star \right]^{1/2}/2$, so
the scalar's decay constant must be larger than
$10^{-2} \sqrt{\kincoup} (H_\star/10^{-22}~\eV)^{-1/4} \Mpl$ to avoid backreaction.

At such large decay constants, the scalar could have an observable impact on cosmology.
In the radiation era, the scalar effectively increases the Hubble rate as small-scale CMB modes
enter the horizon, enhancing diffusion damping of the photon-baryon plasma~\cite{Hou:2011ec,
Bashinsky:2003tk, Baumann:2015rya}.
Bounds on extra radiation content from the CMB largely derive from this effect and currently amount
to a bound $\Delta N_\mathrm{eff} / N_\mathrm{eff} \lesssim 5\%$ to $10\%$.\footnote{
    Runaway scalars redshift like matter after matter-radiation equality, however, providing
    enhanced and distinctive phenomenology compared to pure radiation~\cite{Ramadan:2023ivw,
    Copeland:2023zqz}.
}
Current measurements therefore already limit $f$ to be (roughly) below $\Mpl / 10$, while CMB-S4,
which projects sensitivity to $\Delta N_\mathrm{eff} / N_\mathrm{eff} \sim 1\%$, would probe yet
smaller decay constants $f / \Mpl \sim 0.03$.

% \section{Signatures}
Resonant production mechanisms in general feature dark matter with a density power spectrum sharply
peaked at order unity on some characteristic scale. In the case of axion or scalar oscillations, the
scalar mass is what sets this characteristic wave
number~\cite{Agrawal:2018vin,Bastero-Gil:2018uel,Co:2018lka,Dror:2018pdh,Co:2021rhi,Adshead:2023qiw}.
On the other hand, the vector mass sets a kinematic barrier below which resonant enhancement is
typically inefficient.
The kinetic coupling suppresses the vector mass during production, allowing for peak scales of order
$m_\phi$ which can be far below the present-day dark photon mass.

Density fluctuations at the power spectrum peak collapse shortly after matter-radiation equality,
forming dense small scale structure at astrophysically relevant scales~\cite{Blinov:2021axd}.
The typical peak wave number is set by the Hubble rate at
production~\cite{Cyncynates:2024yxm} as
\begin{align}
    \frac{k_\star}{m_\phi}
    &\approx \sqrt{(\kincoup^2 + \kincoup/2) \frac{2 H_\star}{m_\phi}}
        \left( \sqrt{\kincoup^2 + \kincoup/2} \frac{2 H_\star}{\mA} \right)^{-\frac{1}{2\kincoup + 2}}.
    \label{eqn:kpeak}
\end{align}
In the large-$\kincoup$ limit, the collapse of these density perturbations at matter-radiation equality results in structures with mass
\begin{align}
    M_s
    &\approx 2.4 \times 10^5 M_\odot
        \left( \frac{\kincoup}{10} \right)^{-3}
        \left( \frac{H_\star}{10^{-22}~\eV} \right)^{-3/2}
\end{align}
and radius
\begin{align}
    R_s
    &\approx 1.1~\mathrm{pc} \left( \frac{\kincoup}{10} \right)^{-1}
        \left( \frac{H_\star}{10^{-22}~\eV} \right)^{-1/2}
    .
\end{align}
That minihalos can be much more massive than expected from the dark photon's mass itself provides a
signature that distinguishes this model from other resonant production mechanisms.
While the presence of such massive substructure does not guarantee that the dark photon has kinetic
mixing of any particular size, if an experiment measures a kinetic mixing larger than possible for
other nonthermal production mechanisms, substructure in the dark matter halo would necessarily be so
massive.
\Cref{fig:kinetic-parameter-space} illustrates the mass-coupling parameter space for which upcoming
astrometric~\cite{VanTilburg:2018ykj} and photometric~\cite{Arvanitaki:2019rax, Dai:2019lud} surveys
would be able to probe such extremal substructure.
Other potential signatures include gravitational wave diffraction~\cite{Dai:2018enj}.

% \section{Weak gravity conjecture}
The success of the scalar production mechanism effectively relies on the Abelian-Higgs theory being
extremely weakly coupled at early times, which ostensibly would run afoul of so-called weak gravity
conjectures (WGCs)~\cite{Arkani-Hamed:2006emk}.
While the application of the WGC in its various forms to effective field theory and Higgsed gauge
symmetries is subtle~\cite{Saraswat:2016eaz,Heidenreich:2016aqi,Heidenreich:2017sim,Craig:2018yld},
arguments that gravity becomes strongly coupled at a scale $\Lambda_\mathrm{UV} \sim \gD^{1/3} \Mpl$
are considered robust~\cite{Reece:2018zvv,Harlow:2022ich}.
Requiring that gravity remains weakly coupled at the highest energy scale probed by any production scenario therefore places a lower limit on $\gD$, which, in conjunction with upper bounds from defect formation, can be quite constraining.
Requiring the energy scale of inflation to be below $\Lambda_\mathrm{UV}$ limits viable inflationary production
to $\mA \gtrsim 40~\MeV$.

While it is unclear whether WGCs constrain gauge couplings with displaced moduli, were it so they would
effectively constrain the initial condition of the scalar field via the combination $\kincoup \phi_0 / f$.
Achieving the dark matter relic abundance for a given dark photon mass then requires increasing $H_\star$
[via \cref{eqn:relic-abundance}], diminishing the extent to which the scenario evades defect formation.
Combining the WGC's lower bound on $H_\star$ with the upper bound from defect formation
\cref{eqn:vortex-formation-threshold} places an upper bound on the present-day dark gauge coupling.
These bounds are set by conditions in the early Universe, i.e., requiring the cutoff scale of quantum gravity
to exceed the energy scales of Big Bang Nucleosynthesis (BBN), inflation (if measured), or of the SM
plasma during dark photon production.
Written in terms of the maximum Hubble scale $H_\mathrm{max}$, which for BBN is
$\sim 10^{-15}~\mathrm{eV}$,
\begin{align}
    \frac{g_D}{3 \times 10^{-18}}
    &\lesssim
        \left( \frac{\mA}{10^{-15}~\mathrm{eV}} \right)^{25/22}
        \left( \frac{H_\mathrm{max}}{10^{-15}~\mathrm{eV}} \right)^{-9/22}.
    \label{eqn:wgc-bound}
\end{align}
\Cref{eqn:wgc-bound} takes fiducial values $\kincoup = 25$ and $\lambda = 1$ (and is not
particularly sensitive to either).
This bound is weaker than the defect formation bound but would eliminate most (but not all) of the
prospective parameter space if the inflationary Hubble scale is
$H_\mathrm{inf} \sim 10^{14}~\mathrm{GeV}$ [see \cref{fig:kinetic-parameter-space}].

Weak gravity conjectures also motivate ultralight dark photons receiving their mass from the Higgs mechanism rather than
the St\"uckelberg mechanism.
In supersymmetry, St\"uckelberg fields are accompanied by radial degrees of freedom
with mass $\sim \mA / \gD$ just like the Higgs,
a fact conjectured to hold in any theory of quantum gravity~\cite{Reece:2018zvv}.
In this case, inflationary production of dark photons with a St\"uckelberg mass is just as
constrained as that with a Higgs mass (since the radial modes are produced during inflation if they
are too light).
It would be worth understanding whether the radial mode plays an important role in other
production mechanisms as well.
At the very least, the general threshold for backreaction onto the radial mode should be identical
between the two mass mechanisms.

Though dark matter's current phenomenological relevance resides only at low energies and late times,
identifying its fundamental nature offers myriad opportunities to inform high-energy physics.
Understanding the mechanisms underlying its mass and its early-Universe production reshapes the
implications of direct detection of dark photons.
We explore these implications more broadly in companion work~\cite{Cyncynates:2024yxm}, with this
letter highlighting a scenario whose nonminimality, aside from enabling direct detection in the
laboratory, offers promising and complementary observational signatures.
These results also motivate searching for signatures of dark photon dark matter from purely
gravitational interactions, especially in the near-fuzzy regime
$10^{-22}~\mathrm{eV} \lesssim \mA \lesssim 10^{-15}~\mathrm{eV}$~\cite{Hu:2000ke, Schive:2014dra,
Hui:2016ltb, Hui:2021tkt, Dalal:2022rmp, Amin:2022nlh} where phenomenology can depend on the spin of
dark matter~\cite{Adshead:2021kvl, Zhang:2021xxa, Jain:2021pnk, Gorghetto:2022sue, Amin:2022pzv,
Amin:2022nlh, Jain:2023ojg, Jain:2023tsr, Zhang:2024bjo, Liu:2024pjg, Ling:2024qfv, An:2024axz}.
Further investigation of consistent dark photon cosmologies will deepen our understanding of the
theoretical motivation for and implications of ultralight dark matter detection.

\acknowledgments{
We thank Peter Adshead, Benoit Assi, Masha Baryakhtar, Adrienne Erickcek, Isabel Garcia Garcia,
Anson Hook, Junwu Huang, Justin Kaidi, Justin Khoury, and Mark Trodden for insightful discussions.
D.C. and Z.J.W. are supported through the Department of Physics and College of Arts and Science at
the University of Washington.
This material is partially supported by a grant from the Simons Foundation and the hospitality of
the Aspen Center for Physics.
This work was completed in part at the Perimeter Institute. Research at Perimeter Institute is
supported in part by the Government of Canada through the Department of Innovation, Science and
Economic Development Canada and by the Province of Ontario through the Ministry of Colleges and
Universities.
The dark photon parameter space limits and projections quoted above are compiled in
Ref.~\cite{Caputo:2021eaa, AxionLimits}.
}

\bibliography{bibliography.bib}

\end{document}